# Quantum oscillations in Noncentrosymmetric Weyl semimetals RAlSi (R = Sm, Ce)

Weizheng Cao[1], Qi Wang[1,2], Cuiying Pei[1], Lingling Gao[1], Yi Zhao[1], Changhua Li[1], Na Yu[1], Jinghui Wang[1,2*], Yulin Chen[1,2,3], Jun Li[1,2] and Yanpeng Qi[1,2*]

1. School of Physical Science and Technology, ShanghaiTech University, Shanghai 201210, China
2. ShanghaiTech Laboratory for Topological Physics, ShanghaiTech University, Shanghai 201210, China
3. Department of Physics, Clarendon Laboratory, University of Oxford, Parks Road, Oxford OX1 3PU, UK

* Correspondence should be addressed to Y.Q. (qiyp@shanghaitech.edu.cn) or J.W. (wangjh2@shanghaitech.edu.cn)

## ABSTRACT

Weyl semimetal (WSM) as a new type of quantum state of matter hosting low energy relativistic quasiparticles, has attracted significant attention for both scientific community and potential quantum device applications. Here, we report a comprehensive investigation of the structural, magnetic and transport properties of noncentrosymmetric RAlSi (R = Sm, Ce), which have been predicted to be new magnetic WSM candidates. Both samples exhibit non-saturated magnetoresistance (MR), with ~ 900% for SmAlSi and 80% for CeAlSi at 1.8 K, 9 T. The carrier densities of SmAlSi and CeAlSi display remarkable change around magnetic transition temperatures, signifying that the electronic states are sensitive to magnetic ordering of rare earth elements. At low temperatures, SmAlSi reveals prominent Shubnikov-de Haas (SdH) oscillations associated with the nontrivial Berry phase. High pressure experiments demonstrate that the magnetic order is robust and survival under high pressure. Our results would yield valuable insights of WSM physics and potentials in application to the next-generation spintronic devices in RAX family.

# INTRODUCTION

As a new class of quantum materials, Dirac semimetals (DSMs) and Weyl semimetals (WSMs) have attracted extensive research [1]. The Dirac fermion is protected by time-reversal symmetry (TR) and space-inversion symmetry (SI) [2]. Once the TR or SI is broken, the Dirac point splits into two Weyl points connected by "Fermi arc", one of which starts at the source (+chirality) and the other terminates at the drains (-chirality) [3, 4]. The former option is realized structurally in TaAs and related nonmagnetic materials without SI, while the latter named as magnetic WSMs have been verified experimentally in a handful of materials, such as $Mn_3X$ (X = Sn, Ge), $Co_3Sn_2S_2$, and $Co_2MnGa$ [5-14]. Compared with nonmagnetic ones, magnetic WSMs can exhibit unique quantum transport phenomena, such as the large intrinsic anomalous Hall conductivity (AHC) originating from nontrivial band topology [15]. In this context, magnetic Weyl semimetals are special and may lead to potential applications in spintronics.

Recently, RAlX (R=La, Ce, Pr; X=Si and Ge) with non-centrosymmetric structure have been predicted to host various WSM states and the topological characteristics of Weyl fermions have been detected by angle-resolved photoemission spectroscopy (ARPES) and first-principles calculations [16-18]. This new topological family covers nonmagnetic WSM (R = La) and magnetic WSMs (R = Ce, Pr) depending on the rare-earth elements while keeping the same crystal structure. It also offers remarkable tunability of electronic topology with different magnetic ground states and anisotropic magnetic behaviors by varying rare-earth ions. More importantly, the magnetic members of RAlX present rare examples with breaking both the SI and TR symmetries, thus offering peculiar opportunities to study the interplay between magnetism and Weyl fermions in such an interesting system. Although comprehensive studies on the magnetic RAlX including PrAlGe, CeAlGe, and NdAlSi unveiled some very likely signatures of the Weyl states, experimental identification for existence of Weyl fermions are still insufficient and under debate [16-31]. It definitely needs to be further verified in other RAlX members.

Experimentally, quantum transport study is an important approach to identify the WSM state. In this work, we chose two RAlX members with different magnetic ground states: antiferromagnetic (AFM) SmAlSi and ferromagnetic (FM) CeAlSi. We systematically investigate magnetotransport properties on high-quality single crystals of noncentrosymmetric Weyl semimetals SmAlSi and CeAlSi. Both samples exhibit non-saturated magnetoresistance (MR), ~900% for SmAlSi and 80% for CeAlSi (at 1.8 K, 9 T). The analysis of SdH oscillations reveals two fundamental frequencies originated from the Fermi surface (FS) pockets with non-trivial $\pi$ Berry phases. Our high pressure experiments demonstrate that the magnetic order is robust until high pressure, indicating potential applications in next-generation spintronic devices.

# EXPERIMENTAL DETAILS

The single crystals of RAlSi (R = Sm, Ce) were grown via self-flux method as described elsewhere [19, 32]. High-purity blocks of Samarium/Cerium, Silicon and Aluminum were mixed in the molar ratio of 1 : 1 : 10 and loaded into an alumina crucible. All treatments were performed in an argon-filled glove box, and then the crucible was sealed in a quartz tube under vacuum. The quartz tube was heated up to 1100 °C in 24 h with temperature holding for 12 h in order to ensure guarantee of the raw material melting. Subsequently, the temperature is slowly cooled down to 750 °C at a rate of 2 °C/h and excess Al was removed by a high-speed centrifuge.

The crystal surface morphology and composition were examined by scanning electron microscopy (SEM) and energy dispersive x-ray (EDX) analysis. The phase and quality examinations of both samples were performed on the Bruker AXS D8 Advance powder crystal x-ray diffractometer with Cu $K_{\alpha 1}$ ($\lambda$ = 1.541 78 Å) at room temperature. Rietveld refinements of the PXRD patterns were performed using the TOPAS code [33]. Magnetotransport measurements were performed on Physical Property Measurement System (PPMS). The magnetization measurement was carried on a Magnetic Property Measurement System (MPMS). High-pressure resistivity measurements were performed in a nonmagnetic diamond anvil cell. A cubic BN/epoxy mixture was used for the insulating gaskets and Pt foil was employed in the electrical leads. Pressure was determined by the ruby luminescence method [34].

**RESULTS AND DISCUSISION**

Figure 1(a) depicts the schematic structure of RAlGe (R = Sm, Ce) crystallized in the tetragonal LaPtSi-type structure with space group $I4_1md$ (No. 109). The structure consists of stacks of rare-earth elements (R), Al, and Si layers, and along the [0 0 1] direction each layer consists of only one type of element. Chemical composition analysis using EDX demonstrated that the one-to-one correspondence between actual and nominal compositions is good [Fig. 1(b)]. The refinement powder XRD patterns of SmAlSi and CeAlSi are shown in Figure 1(c) and (d), respectively. The Bragg reflections can be well refined by the Rietveld method with reliability parameters, manifesting high-quality RAlGe (R = Sm, Ce) crystallized into the noncentrosymmetric structure. The lattice parameters of RAlSi (R = Sm, Ce) obtained in this work are summarized in Table I.

Figure 2(a) shows the temperature ($T$) dependence of longitudinal resistivity ($\rho$) for SmAlSi. Under zero field, SmAlSi exhibits a typical metallic behavior with residual resistivity of $\rho_0$ = 17.62 μΩ cm and residual resistivity ratio RRR = 4.95. The resistivity of SmAlSi shows a small anomaly at $T_N$ ~ 10.7 K, as shown in the inset of Figure 2(a), corresponding to AFM transition [35]. This magnetic transition was further confirmed by the magnetic susceptibility and isothermal magnetization measurements, shown in Figure 2(b) and 2(c). The effective moment estimated from the Curie-Weiss fit of the magnetic susceptibility above 10 K results in $\mu_{\text{eff}}$ ~ 1.19 $\mu_B$, close to the theoretical one of $\mu_{\text{eff}}^{\text{Sm }3+}$ ~ 0.84 $\mu_B$. We also performed the electrical transport and magnetization measurements on high-quality single crystals of CeAlSi. Different from SmAlSi, CeAlSi hosts a FM order with Curie temperature ($T_C$) ~ 9.4 K and effective magnetic moment $\mu_{\text{eff}}$ ~ 3.51 $\mu_B$/Ce.

Figure 3(a) shows temperature dependent magnetoresistance (MR) of SmAlSi, where the magnetic field is along the [0 0 1] direction. The MR is defined as MR = $[\rho_{xx}(B) - \rho_{xx}(0)]/\rho_{xx}(0) \times 100\%$ in which $\rho_{xx}(B)$ and $\rho_{xx}(0)$ represent the resistivity with and without magnetic field, respectively. At low temperatures, SmAlSi exhibits a large non-saturated MR behavior and MR reaches ~ 900% under 9 T at 1.8 K. Figure 3(b) shows the Hall resistivity $\rho_{yx}(B)$ at different temperatures. At 2 K, $\rho_{yx}(B)$ shows a nonlinear behavior and its slope changes sign from positive at low fields to negative at high fields, indicating the two-type carriers coexistent in SmAlSi. To clarify two types of carriers with temperature, we fitted the data using the two-band model [36]:

$$\rho_{xx} = \frac{1}{e} \frac{(n_e\mu_e + n_h\mu_h) + \mu_e\mu_h(n_e\mu_h + n_h\mu_e)B^2}{(n_e\mu_e + n_h\mu_h)^2 + \mu_e^2\mu_h^2(n_h - n_e)^2 B^2}$$

$$\rho_{yx} = \frac{B}{e} \frac{(n_h\mu_h^2 - n_e\mu_e^2) + \mu_e^2\mu_h^2(n_h - n_e)B^2}{(n_e\mu_e + n_h\mu_h)^2 + \mu_e^2\mu_h^2(n_h - n_e)^2 B^2}$$

where $\mu_e$, $\mu_h$ are the mobilities of electrons and holes, and $n_e$, $n_h$ are the concentrations of electrons and holes, respectively. Figure 3(c) shows carrier concentrations and mobilities as a function of temperature. The carrier densities at 1.8 K reach $n_e = 10.88 \times 10^{25}$ m$^{-3}$ and $n_h = 8.53 \times 10^{25}$ m$^{-3}$ with carrier mobilities $\mu_e = 1.76 \times 10^3$ cm$^2 \cdot$ V$^{-1} \cdot$ s$^{-1}$ and $\mu_h = 4.20 \times 10^3$ cm$^2 \cdot$ V$^{-1} \cdot$s$^{-1}$, which is comparable with other topological semimetals. It should be noted that the carrier densities ($n_e$ and $n_h$) display a kink around $T_N$, indicating that the electronic states are sensitive to magnetic ordering of Sm moments.

Since the MR of SmAlSi at 1.8 K and 9 T only displays very weak quantum oscillation that is rather difficult for further analysis, the magnetotransport measurements were then subjected to lower temperature and higher magnetic field. Figure 4(a) presents the resistivity dependence of magnetic field at various temperatures with field up to 14 T. Striking Shubnikov-de Haas (SdH) quantum oscillations in the MR are visible. After subtracting the smooth background, the SdH oscillations at different temperatures from 2 to 50 K against the reciprocal magnetic field 1 / B are presented in Figure 4(b). The oscillation patterns show obviously multi-frequency behavior. From the Fast Fourier Transform analysis (FFT) of SdH oscillations, two fundamental frequencies $F_\alpha$ = 11.8 T and $F_\beta$ = 35.3 T with their harmonic frequencies $F_{2\beta}$ = 70.5 T are clearly identified, indicating the presence of at least two Fermi surface pockets at the Fermi level, as shown in Figure 4(c). According to the Onsager relation $F = (\phi_0/2\pi^2)A_F = (\hbar/2\pi\, e)A_F$, the cross-sectional area of the Fermi surface normal to the field are $A_\alpha$ = 1.1×10$^{-3}$ Å$^{-2}$ and $A_\beta$ =3.4×10$^{-3}$ Å$^{-2}$. The values of $A_F$ are comparable with recently report [31].

The Dirac/Weyl system will produce a nontrivial $\phi_B$ under a magnetic field, which could be probed by using the Landau level (LL) index fan diagram or a direct fit to the SdH oscillations by using the Lifshitz-Kosevich (LK) formula. To gain insight into the topological states of SmAlSi, we perform the analysis of quantum SdH oscillations

using LK formula [37-39]:

$$\Delta\rho_{xx}^i \propto \frac{5}{2}\sqrt{\frac{B}{2F}} R_T R_D R_S \cos[2\pi(F/B + \gamma - \delta + \varphi)]$$

where $R_T = \lambda\mu T / B \sinh(\lambda\mu T / B)$, $R_D = \exp(-\lambda\mu T_D / B)$, $R_S = \cos(\pi\mu g^*)$. Here, $\mu = m^* / m_0$ is the ratio of effective cyclotron mass $m^*$ to free electron mass $m_0$, $T_D$ is the Dingle temperature, $g^*$ is the effective g factor, and $\lambda = 2\pi^2 k_B m_0/e\hbar \approx 14.7$ T/K [40]. The oscillatory components of $\Delta\rho$ is expressed by the cosine term with a phase factor $\gamma - \delta + \varphi$, in which $\gamma = 1/2 - \phi_B/2\pi$, $\phi_B$ is Berry phase. The small effective cyclotron mass $m^*$ at the $E_F$ could be obtained by fitting the temperature dependence of the FFT magnitude by the temperature damping factor $R_T$, as shown in Figure 4(d), giving $m_\alpha^* = 0.10\ m_0$ and $m_\beta^* = 0.09\ m_0$ [41-43]. Additionally, the obtained Dingle temperature $T_D$ are 7.9 K and 19.2 K for $F_\alpha$ and $F_\beta$, respectively. As shown in the Figure 4(e), the LK formula reproduces the resistivity oscillations well at 0.7 K. Taking $\delta = \pm 1/8$ for three-dimensional system and $\varphi = 1/2$ for $\rho_{xx} \gg \rho_{xy}$, the yielded Berry phases $\phi_B$ are $(1.07 \pm 0.25)\pi$ for $F_\alpha$, and $(1.32 \pm 0.25)\pi$ for $F_\beta$, respectively. Both Fermi pockets exhibit nontrivial Berry phases [44-46]. The results are summarized in Table II. We present the Landau level (LL) fan diagram shown in figure 4(f). All the points almost fall on a straight line, thus allowing a linear fit that gives an intercept of ~ 0.965 and 0.946 for $\alpha$ and $\beta$ pocket, respectively. This further demonstrates the nontrivial topological states in SmAlSi.

We also preform magnetotransport measurements for CeAlSi. The MR and $\rho_{yx}$ of CeAlSi as a function of magnetic field, with $B$ along the [0 0 1] direction, are also shown in Figures 3(d) and 3(e), respectively. MR increases with magnetic field without the trend of saturation. The magnitude of Hall resistivity $\rho_{yx}$ increases down to 100 K. Since CeAlSi is a typical ferromagnetic with $T_C \sim 9.4$ K, a small AHE is observed. The Hall resistivity $\rho_{yx}$ in a ferromagnet traditionally has two parts [47, 48]:

$$\rho_{yx} = \rho_{yx}^0 + \rho_{yx}^A = R_0 B + 4\pi R_s M$$

where $\rho_{yx}^0$ is ordinary Hall resistivity, $\rho_{yx}^A$ is the anomalous Hall resistivity, $R_0$ represents the ordinary Hall coefficient and $R_s$ represents anomalous Hall coefficient. Here, $\rho_{yx}$ exhibits almost linear behavior below 100 K. We obtain the $R_0$ by fitting the data at the magnetic field range of 6 - 9 T, which $R_0$ has the negative value indicating the major role of electron-type carriers for the transport. The electron carrier concentration and the mobility dependence temperature were obtained and shown in Figure 3(f). We also noted that the electron carriers display a kink around $T_C$ indicating that the electronic states are sensitive to magnetic ordering of Ce moments. Unfortunately, the low-field MR of CeAlSi displays no quantum oscillations, the magnetotransport measurements are thus needed to subject to high magnetic field.

At ambient pressure, both samples have magnetic ground states at low temperature, we then investigate the effect of pressure on the magnetic properties for RAlGe (R = Sm, Ce). Figures 5(a) and 5(b) show the plots of temperature versus resistivity under various pressures for SmAlSi and CeAlSi, respectively. Both samples show a metallic behavior in the whole pressure range. Application of pressure has weak effect on the magnetic ground state for both SmAlSi and CeAlSi. As shown in Figures 5(c) and (d), the Néel temperature $T_N$ and Weiss temperature $T_C$ is survival up to 46.2 GPa and 21.4 GPa for

SmAlSi and CeAlSi, respectively. No superconductivity was observed down to 1.8 K in this pressure range [49, 50]. Magnetic Weyl semimetal RAlGe (R = Sm, Ce) with nontrivial topology of electronic states display robust magnetic ground states upon compression, which have potential applications in the next-generation spintronic devices.

**CONCLUSION**

In conclusion, we synthesized high quality single crystals and performed comprehensive magnetotransport studies on magnetic Weyl semimetal RAlSi (R = Sm, Ce). Both samples exhibit non-saturated MR and robust magnetic order even under high pressure. The analysis of SdH oscillations of SmAlSi reveals two fundamental frequencies originated from the FS pockets with non-trivial $\pi$ Berry phases. Considering both the SI and TR symmetries breaking, our results call for further experimental and theoretical studies on RAlSi family and related materials for a better understanding of the interplay between magnetic and topological nature, and its potential application in realizing spintronic devices.

**ACKNOWLEDGMENT**

This work was supported by the National Key R&D Program of China (Grants No. 2018YFA0704300, 2017YFB0503302), the National Natural Science Foundation of China (Grant No. U1932217, 11974246, 12004252, 61771234, 12004251), the Natural Science Foundation of Shanghai (Grant No. 19ZR1477300, 20ZR1436100), the Science and Technology Commission of Shanghai Municipality (19JC1413900, YDZX20203100001438), the Shanghai Sailing Program (Grant No. 21YF1429200), the Interdisciplinary program of Wuhan National High Magnetic Field Center (WHMFC202124), and the Beijing National Laboratory for Condensed Matter Physics. The authors thank the support from Analytical Instrumentation Center (# SPST-AIC10112914), SPST, ShanghaiTech University.

TABLE I. Structure parameters of RAlSi (R = Sm, Ce)

| Compound | SmAlSi | CeAlSi |
|---|---|---|
| Space group | $I4_1md$ (No. 109) | |
| Crystal structure | LaPtSi-type | |
| $a$ | 4.1583 Å | 4.2550 Å |
| $c$ | 14.4562 Å | 14.5919 Å |
| $V$ | 249.9725 Å$^3$ | 264.1857 Å$^3$ |
| $Z$ | 4 | 4 |
| $R_p$ | 4.91 % | 4.67 % |
| $R_{wp}$ | 8.01 % | 7.32 % |

TABLE II. Parameters extracted from SdH oscillation for SmAlSi

| $F$ (T) | $A_F(10^3\text{Å}^{-2})$ | $k_F(10^{-2}\text{Å}^{-2})$ | $m^*(m_0)$ | $v_F(10^5 \text{ms}^{-1})$ | $E_F$(meV) | $T_D$(K) | $\tau_D(10^{-13}\text{s})$ | $\phi_B$ | $\mu$ (m$^2$ V$^{-1}$ s$^{-1}$) |
|---|---|---|---|---|---|---|---|---|---|
| 11.8 | 1.1 | 1.9 | 0.1 | 2.2 | 27.5 | 7.9 | 1.03 | $(1.07 \pm 0.25)\pi$ | 0.18 |
| 35.3 | 3.4 | 3.3 | 0.09 | 4.2 | 90.2 | 19.2 | 0.34 | $(1.32 \pm 0.25)\pi$ | 0.07 |

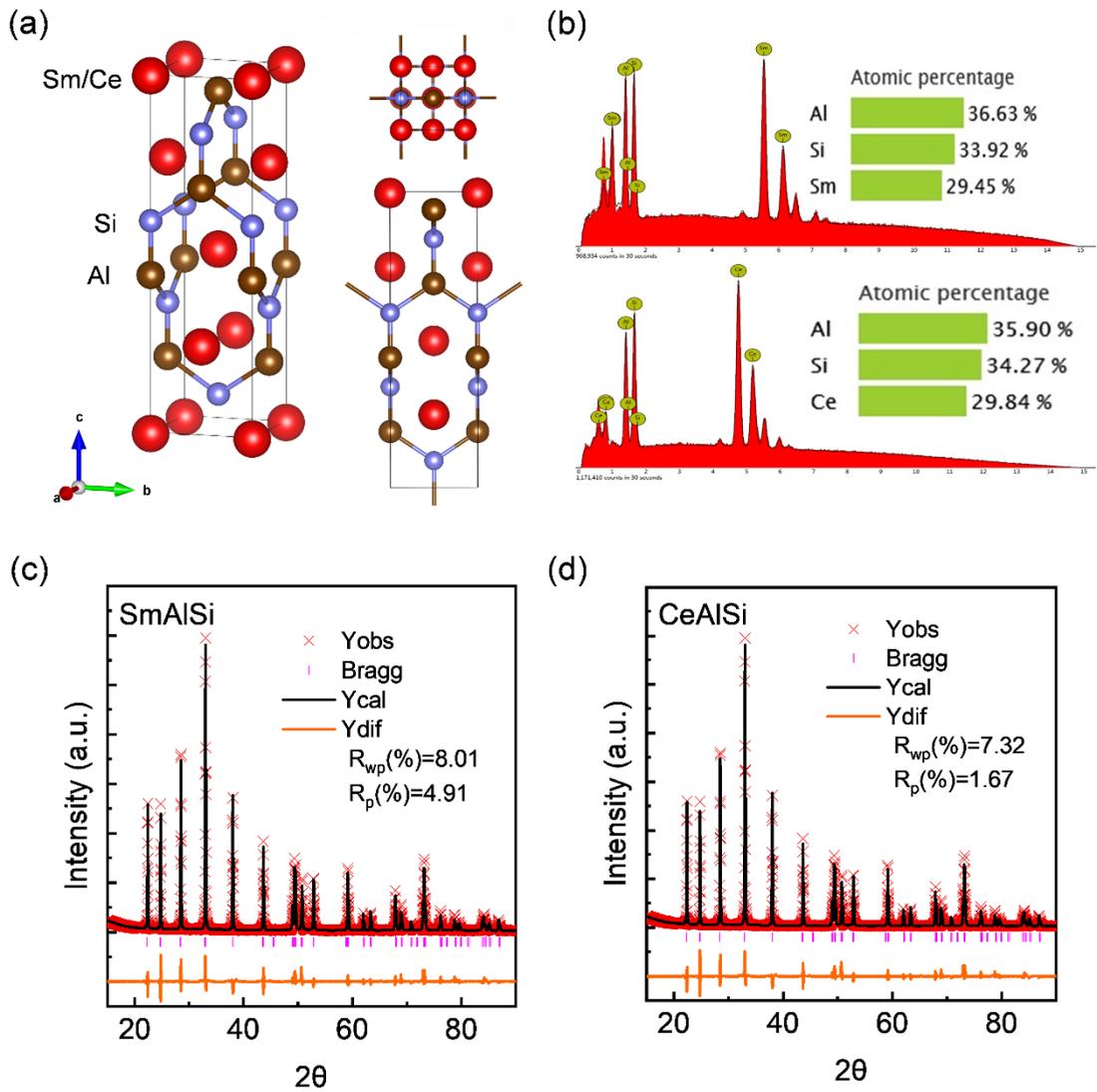

FIG. 1. (a) The crystal structure of RAlSi (R = Sm and Ce). Top and side (as view from *a*-axis) of RAlSi. (b) Typical EDX spectrum of the SmAlSi and CeAlSi single crystals. (c) and (d) Rietveld refinement of the powder x-ray diffraction patterns of SmAlSi and CeAlSi, respectively.

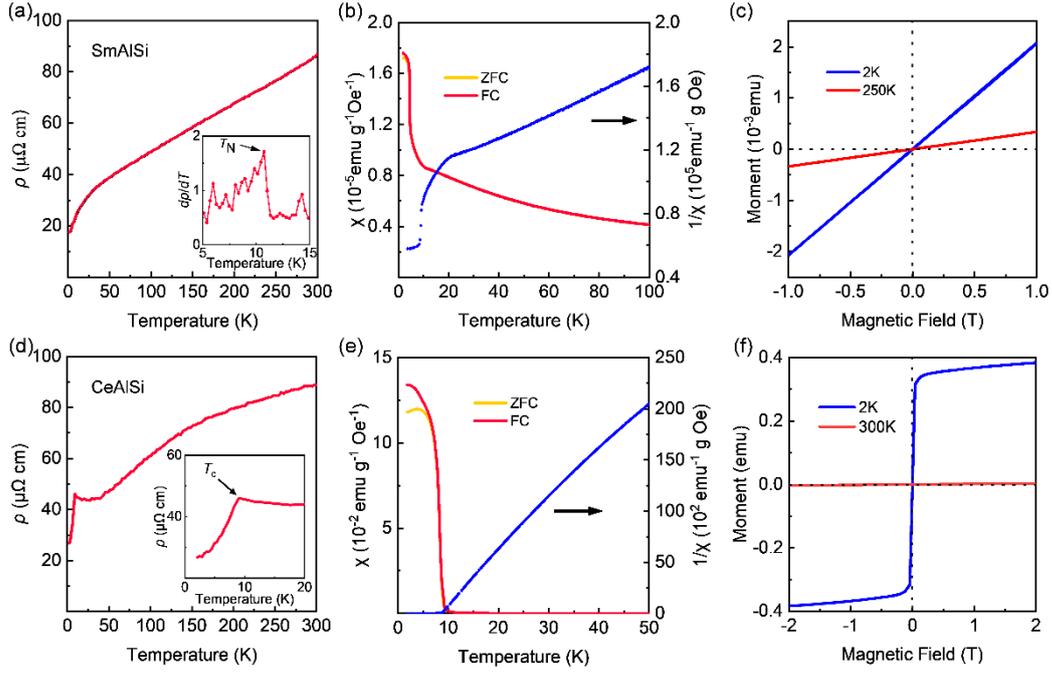

FIG. 2. (a) and (d) Temperature dependence of resistivity of SmAlSi and CeAlSi in the temperature range between 1.8 K and 300 K. Inset of Fig. (a): d$\rho$/d$T$ of SmAlSi. The $T_N$ (~ 10.7 K) is taken from the peck of d$\rho$/d$T$ curve. Inset of Fig. (d): enlarged view of low temperature, showing $T_C$ ~ 9.4 K. (b) and (e) Temperature dependence of the in-plane magnetic susceptibility for SmAlSi and CeAlSi, respectively. The inverse of the magnetic susceptibility is also shown here. (c) and (f) Field dependence of the in-plane magnetic moment for SmAlSi and CeAlSi, respectively.

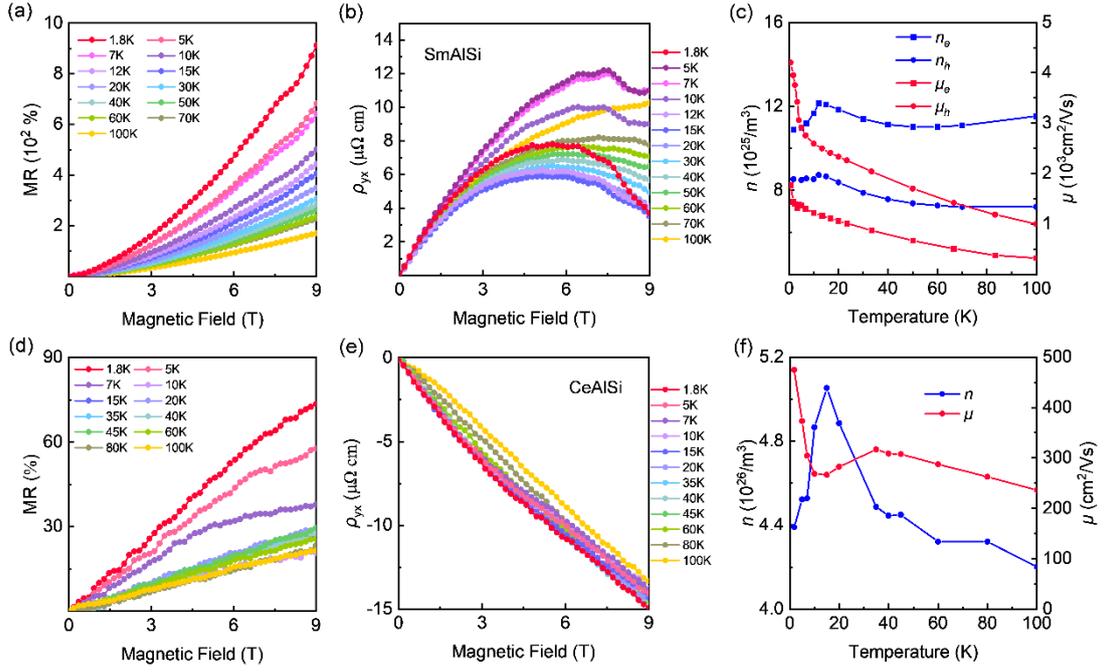

FIG. 3. (a) and (b) Longitudinal resistivity as a function of magnetic field *B* at various temperature for SmAlSi and CeAlSi, respectively. The magnetoresistance (MR) is defined as $MR = [\rho_{xx}(B) - \rho_{xx}(0)]/\rho_{xx}(0) \times 100\%$ in which $\rho(B)$ and $\rho(0)$ represent the resistivity with and without *B*, respectively. (b) and (e) Hall resistivity for SmAlSi and CeAlSi, respectively. (c) and (f) Carrier concentration and mobility of SmAlSi and CeAlSi as a function of temperature, respectively.

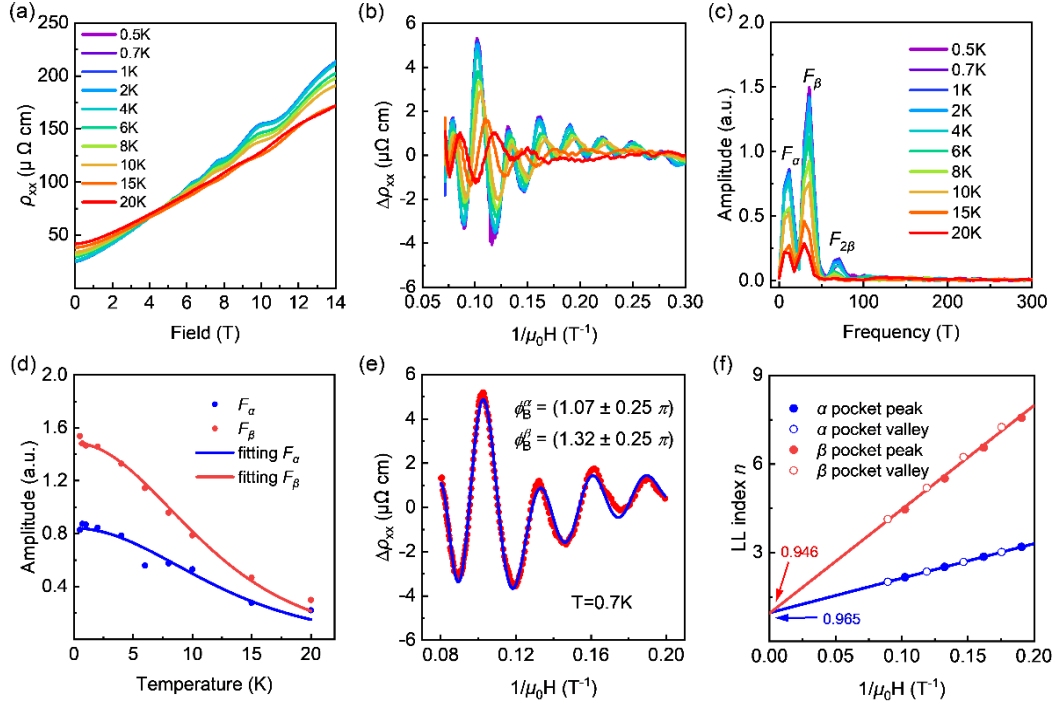

FIG. 4. (a) Magnetic field dependence of resistivity for SmAlSi at various temperatures with the magnetic field along the *c*-axis. (b) Oscillation patterns $\Delta\rho_{xx} = \rho_{xx} - \langle\rho_{xx}\rangle$ plotted as a function of $1/\mu_0 H$ at various temperature. (c) FFT spectra of $\Delta\rho_{xx}$ at different temperatures. Two major frequencies of $F_\alpha = 11.8$ T and $F_\beta = 35.3$ T with their harmonic frequencies $F_{2\beta} = 70.5$ T are extracted. (d) The fits of the FFT amplitudes of $F_\alpha$ and $F_\beta$ to the temperature damping factor $R_T$ of the LK formula. (e) The LK fits (blue lines) of SdH oscillations at 0.7 K. (f) Landau level (LL) fan diagram for $F_\alpha$ and $F_\beta$ at 0.7 K. The solid lines are the linear fits to the experimental data. The values of the intercepts of the fitting lines with the LL index axis are shown here.

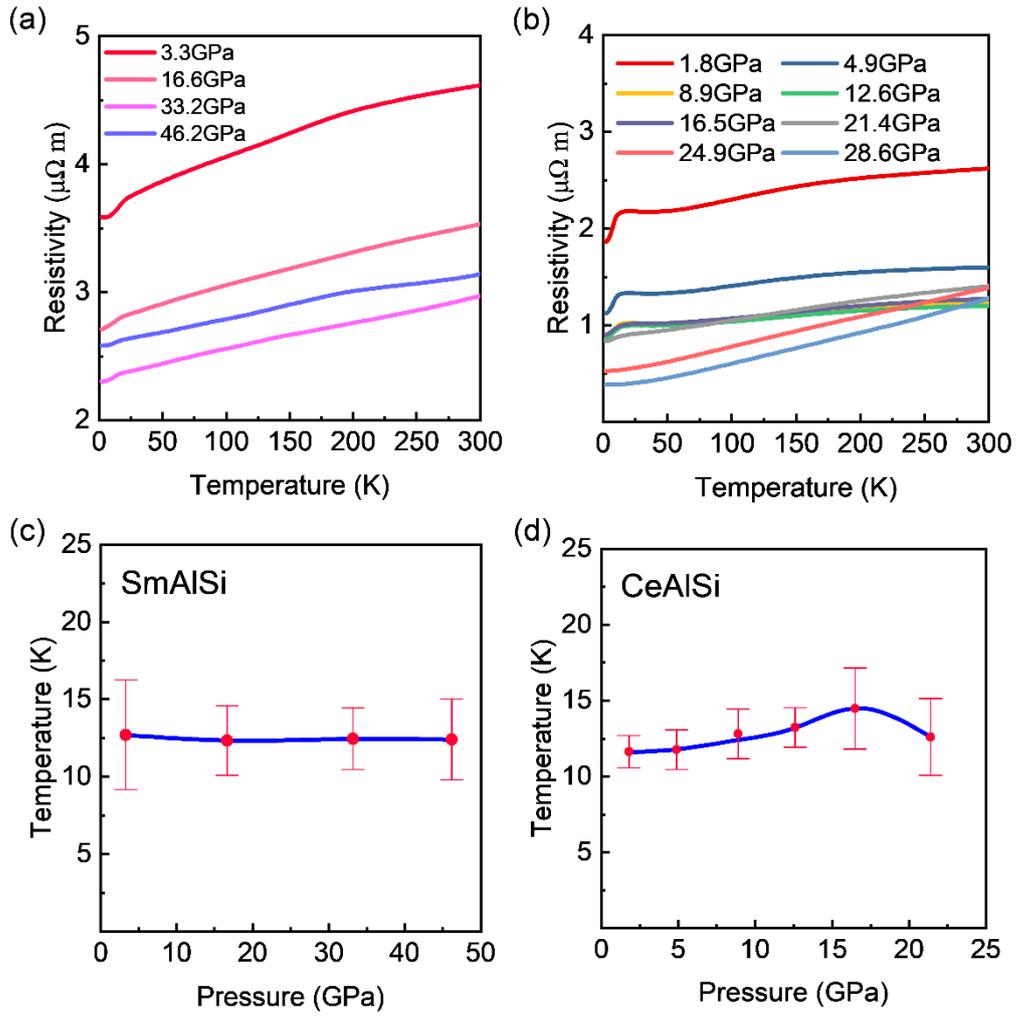

FIG. 5. (a) and (b) Temperature dependence of electrical resistivity at various pressures for SmAlSi and CeAlSi, respectively. (c) and (d) Pressure dependence of magnetic transition temperature for SmAlSi and CeAlSi, respectively.